\documentstyle[aps,amsmath,amssymb]{revtex}  

\begin{document}

\def\i{{\rm i}\,}
\newcommand{\bigfrac}[2]{\mbox {${\displaystyle \frac{ #1 }{ #2 }}$}}
\newcommand{\bra}[1]{\left\langle #1 \right |}
\newcommand{\ket}[1]{\left | #1 \right\rangle}
\newcommand{\expect}[1]{\left\langle #1 \right\rangle}
\newcommand{\beq}{\begin{equation}}
\newcommand{\beqa}{\begin{eqnarray}}
\newcommand{\eeq}{\end{equation}}
\newcommand{\eeqa}{\end{eqnarray}}
\newenvironment{eqblock}[2]{\beq\label{#2}\begin{array}{#1}}{\end{array}
                                \eeq}
\newenvironment{neqblock}[1]{\[\begin{array}{#1}}{\end{array}\]}
\newcommand{\beqb}{\begin{eqblock}}
\newcommand{\eeqb}{\end{eqblock}} 

\title{Out of equilibrium correlation functions of quantum anisotropic 
$XY$ models: one-particle excitations.}
\author{Luigi Amico and Andreas Osterloh}
\address{ NEST-INFM $\&$ Dipartimento di Metodologie Fisiche e 
	Chimiche (DMFCI), viale A. Doria 6, 95125 Catania, ITALY}

\maketitle

\begin{abstract}
We calculate exactly matrix elements between states that are not eigenstates 
of the quantum XY model for general anisotropy. Such quantities therefore 
describe non equilibrium properties of the system; the Hamiltonian
does not contain any time dependence. 
These matrix elements are expressed
as a sum of Pfaffians. For single particle excitations on the ground state 
the Pfaffians in the sum simplify to determinants. 
\end{abstract}

\section{Introduction}

Spin systems are paradigmatic models for describing many phenomena in 
contemporary physics\cite{AUERBACH}. 
Often their main properties can be captured qualitatively,  
resorting on approximated techniques.
However, in many cases more refined approaches are required to 
obtain reliable results. 
A prime example are systems near or at a phase transition, where 
quantum fluctuations inhibit many standard routes from working 
(as e.g mean-field theory but also perturbation theory).  
At the critical point, in change, the system can remarkably be simplified 
by a then present large class of symmetries. Conformal field theory 
employs systematically this important property of the system at criticality
and the corresponding dynamics can be integrated exactly 
in 1+1 dimensions. 
Much more difficulties arise when the system is far from either, 
criticality and accessibility to mean-field or perturbation theory. 
Fortunately there are many non-trivial systems for which
the symmetry is large enough to allow the dynamics 
being integrated exactly even for generic values of the relevant couplings. 
Then, also the physics of the cross over from non-critical to critical 
regimes is accessible. 
Complete integrability constitutes the crucial property that 
even exact
correlation functions are available. Important steps forward to this goal 
have become possible by the Quantum Inverse Scattering 
approach~\cite{KOREPIN-BOOK}, more recently refined by  Kitanine, Maillet, Slavnov and 
Terras~\cite{INVERSION1} and  Korepin and G\"ohman~\cite{INVERSION2}.

The quantum anisotropic $XY$ chains are a relevant 
example of completely integrable models.
The model was solved exactly by Lieb, Schultz and Mattis~\cite{Lieb61}, 
Pfeuty~\cite{Pfeuty70} for isotropic cases and by 
Barouch, McCoy, and Dresden~\cite{Mccoy70} for generic anisotropy. Also
the correlation functions were intensively studied and
analytic expressions for their asymptotics (in time and space variables) 
were obtained~\cite{Mccoy70,ITS}. 
The correlation functions were calculated at equilibrium and for time-dependent
magnetic field.
We perform an 
exact calculation of correlations between states that are {\it not 
eigenstates} of the model, and that therefore describe  
{\it non-equilibrium} properties of the model; we remark that 
the Hamiltonian instead does not contain explicit time-dependence.     
Our motivations come from condensed matter where
quantum $XY$ chains are particularly studied, even more 
intensively since recent interest in the phenomenon of decoherence in 
suitably designed physical systems~\cite{Khveshchenko03}; 
this latter kind of analysis is due, 
in turn, to the burst of interest in quantum information theory~\cite{Nielsen00}. 
Such cross-over of interests originated a line of research investigating 
the interconnection between condensed matter and 
quantum information. In particular it is 
intriguing  to investigate whether it is  
possible to better characterize condensed matter states by 
looking at e.g. quantum correlations {\it entanglement} properties of their wavefunction.
Already a number of interesting results in this direction have 
been obtained~\cite{Preskill00}-\cite{DYNAMICS}.

The present paper is laid out as follows. In the next section we present the models
we discuss and review the exact solution from Refs.~\cite{Lieb61,Mccoy70},
already preparing relevant building blocks for computing off-equilibrium correlations.
In section \ref{corr-as-pfaffians} we present known results connecting 
vacuum expectation values in fermionic theories with a generalized determinant structure,
called the Pfaffian and their application to equilibrium correlation functions
presented in \cite{Pfeuty70,Mccoy70}. Section \ref{result} contains the
main result for non-equilibrium correlations and matrix elements of the presented models.
After all we draw our conclusions.

\section{The models}\label{models}
The system under consideration is a spin-1/2 ferromagnetic chain with
an exchange coupling $\lambda$ in a transverse 
magnetic field of strength $h$. 
The Hamiltonian is $H=h H_s$ with the dimensionless Hamilton operator
$H_s$ being
\begin{equation}
H_s=-\lambda \sum_{i=1}^N (1+\gamma)S^x_i S^x_{i+1}+
(1-\gamma)S^y_i S^y_{i+1} - \sum_{i=1}^N S^z_i
\label{model} 
\end{equation}
where $S^a$ are the spin-$1/2$ matrices ($a=x,y,z$) and $N$ is 
the number of sites. We assume periodic boundary conditions. 
The anisotropy parameter $\gamma$ connects the quantum Ising model 
for $\gamma =1$ with the isotropic XY model for $\gamma = 0$.
In the interval  $0<\gamma\le 1$ the model belongs to the Ising 
universality class and for $ N =\infty$ it undergoes a quantum phase 
transition at the critical coupling $\lambda_c=1$. The order parameter is
the magnetization in $x$-direction, $\langle 
S^x\rangle $, which is different from zero for $\lambda >1$ and 
vanishes at and below the transition. 
On the contrary the magnetization along the $z$-direction, 
$\langle S^z\rangle $, is different from zero for any value of $\lambda$. 

This class of models was diagonalized by means of the 
Jordan-Wigner transformation~\cite{Lieb61,Pfeuty70,Mccoy70}  that maps spins 
to one dimensional spin-less fermions with creation and annihilation operators
$c^\dagger_l$ and $c^{}_l$. It proved convenient to use the operators
$A_l\doteq c_l^\dagger + c^{}_l$, $B_l\doteq c_l^\dagger - c^{}_l$,
which fulfill the anti-commutation rules 
\begin{eqnarray}
\{A_l, A_m\}&=&-\{B_l, B_m\}=2 \delta_{lm} \; ,\nonumber \\  
\{A_l, B_m\}&=&0 \; .
\label{A-B-algebra}
\end{eqnarray}
In terms of these operators the Jordan-Wigner transformation reads
\begin{eqnarray}
S_l^x&=&\frac{1}{2}A_l \prod_{s=1}^{l-1} A_s B_s \nonumber \\
S_l^y&=&-\frac{i}{2}B_l \prod_{s=1}^{l-1} A_s B_s \nonumber \\
S_l^z&=&-\frac{1}{2} A_l B_l \;.
\label{jordanwigner}
\end{eqnarray}  
The Hamiltonian defined in Eq.(\ref{model}) is bilinear in the 
fermionic degrees of freedom and therefore can be diagonalized 
by means of the transformation
\begin{equation}
\eta_k=\frac{1}{\sqrt{N}}\sum_l e^{ikl}\left (\alpha_k c^{}_l+i \beta_k c_l^\dagger \right ) 
\label{eta}
\end{equation}
with coefficients 
\begin{eqnarray}
\alpha_k &=& \frac{\Lambda_k-(1+\lambda \cos k)}{\sqrt{2 [\Lambda_k^2 
-(1+\lambda \cos k) \Lambda_k ]}} \nonumber \\
\beta_k &=& \frac{\gamma \lambda \sin k}{\sqrt{2 [\Lambda_k^2 
-(1+\lambda \cos k) \Lambda_k ]}} \; .
\end{eqnarray}
The Hamiltonian thereafter assumes the form
\begin{equation}
H=\sum_k \Lambda_k \eta_k^\dagger \eta_k 
- \bigfrac{1}{2}\sum_k \Lambda_k 
\end{equation} 
and the associated energy spectrum is 
$$\
\Lambda_k= \sqrt{ \left (1+\lambda \cos k\right )^2 +
\lambda^2 \gamma^2 \sin^2 k}\; .
$$ 
Now, in order to calculate correlations out of equilibrium, we
need to know the time dependence of the relevant operators.
From Eq.(\ref{eta}) we obtain the spin-less fermion creation and annihilation
operators in the Heisenberg picture.
We have $\eta^\dagger_k(t)=\exp{(-\i \Lambda_k t)}\eta^\dagger_k(0)$
and hence, using Eq.(\ref{eta}) and its inverse
$$
c^{}_j(t)= \sum_l [\tilde{a}_{l-j}(t) c^{}_l - \tilde{b}_{l-j}(t)c^\dagger_l ]
$$
where the new coefficients are
\begin{eqnarray}
\tilde{a}_{x}(t) &=& \frac{1}{\sqrt{N}}\sum_{k} \cos{k x} 
	\left ( e^{i \Lambda_k t}-2i \beta^2_k \sin \Lambda_k t\right ) \\
\tilde{b}_{x}(t) &=& \frac{2\i}{\sqrt{N}}\sum_{k} \sin{k x}\,
	\alpha_k \beta_k \sin \Lambda_k t \;\; .
\end{eqnarray}

In the limit $\gamma =0$ the previous expressions simplify 
considerably.
In this case the magnetization, i.e. the $z$-component of the total spin 
$S^z=\sum_j S^z_j$, is a conserved quantity. In terms of
fermions this corresponds to the conservation of the total number 
of particles, $N=\sum_j n_j=\sum_j c^\dagger_j c^{}_j$.
For $\gamma\longrightarrow 0$  and $|\lambda|\leq 1$ we find that  
$\alpha_k\longrightarrow 0$ and $\beta_k\longrightarrow {\rm sign}\,k$.
The energy spectrum is $\Lambda_k=\left|1+\lambda \cos k\right|$ and
the eigenstates are plane waves 
(the Hamiltonian corresponds to a tight binding model)
\beq
\label{c(t):gamma0}
c^{}_j(t) = \frac{1}{\sqrt{N}}\sum_{k}\sum_{l} \cos{k(l-j)} 
e^{-i \Lambda_k t} c^{}_l 
\eeq
\begin{equation}
\eta^\dagger_k=\frac{1}{\sqrt{N}}\sum_l e^{-ikl} c^{}_l\; .
\label{eta:gamma0}
\end{equation}
In this work we discuss vacuum expectation values and 
correlations in excitations of them.
It is worthwhile noticing that different strategies are applied, depending
on whether the vacuum is the ground state or the state with no 
particles (which we call the  $c$-vacuum). 
Since it is cumbersome to calculate the time dependence of the vacuum itself,
it is convenient to write the operators $A_l$ and $B_l$ in the 
Heisenberg picture.
For the ground state instead the time dependence is trivial and
the operators are taken in the Schr\"odinger picture.
For both approaches we express the operators $A_l$ and $B_l$ in terms of the
operators $\eta_k$ and $\eta_k^\dagger$
\beqa\label{A-of-eta}
A_l &=& \bigfrac{1}{\sqrt{N}}\sum_q\left[\eta^\dagger_{-q}+\eta^{}_q\right]
	z_q e^{-\i q l} \\
\label{B-of-eta}
B_l &=& \bigfrac{1}{\sqrt{N}}\sum_q\left[\eta^\dagger_{-q}-\eta^{}_q\right]
	z^*_q e^{-\i q l}
\eeqa
These are sufficient for the correlations in the ground state.
For the calculation for the $c$-vacuum it proves to be convenient 
defining the following (redundant) Fourier transforms 
containing $\alpha_k$, $\beta_k$, and
their combination $z_k:=\alpha_k + \i \beta_k$
\beqa
{\frak A}_x&:=&\bigfrac{1}{L}\sum_q {\rm d}q 
	\alpha^2_q e^{\i q x}\\
{\frak B}_x&:=&\bigfrac{1}{L}\sum_q {\rm d}q 
	\beta^2_q e^{\i q x}\\
{\frak Z}_x&:=&\bigfrac{1}{L}\sum_q {\rm d}q 
	z^2_q e^{\i q x}\\
{\frak AZ}_x(t)&:=&\bigfrac{1}{L}\sum_q {\rm d}q 
	\alpha_q z_q e^{\i \Lambda_q t} e^{\i q x}\\
{\frak BZ}_x(t)&:=&\bigfrac{1}{L}\sum_q {\rm d}q 
	\beta_q z_q e^{\i \Lambda_q t} e^{\i q x}\\
\mu_x(t)&:=&{\frak AZ}^*_x(t)-\i {\frak BZ}_x(t)\; .
\eeqa
In these quantities we have
\beqa\label{A-of-t}
A_l(t)&=&\sum_j \left( c_j^\dagger \mu_{j-l}(t) + c^{}_j \mu^*_{j-l}(t)\right)\\
\label{B-of-t}
B_l(t)&=&\sum_j \left( c_j^\dagger \mu_{l-j}(t) - c^{}_j \mu^*_{l-j}(t)\right)\; .
\eeqa
where
\beq
\mu_x(t)=\bigfrac{1}{L}\sum_q {\rm d}q e^{-\i\Lambda_q t}
	(\alpha_q^2 \cos qx + \alpha_q \beta_q \sin qx)\; .
\eeq

\section{Correlation functions as Pfaffians}\label{corr-as-pfaffians}
It is known since 1952 that ``vacuum'' expectation values of a product
of $2 R$ fermionic fields 
\beq\label{vacuum-expect}
\bra{0}\Psi_1\cdots\Psi_{2R}\ket{0}
\eeq
can be written as a {\em Pfaffian}\cite{Caianello52}.
The entries of the Pfaffian structure are the {\em contractions }
of two field operators
\beq\label{contraction}
\bra{0}\Psi_i\Psi_j\ket{0}=P_{i,j}.
\eeq
The field operators $\Psi$ are linear functionals of fermionic 
creation and annihilation operators, where the 
``vacuum'' $\ket{0}$ is that state annihilated by the annihilation operators. 
The Pfaffian is a type of generalized determinant form\cite{Caianello52}. 
It is written in a triangular structure as
$$
\sum\limits_{\pi\in {\cal S}^{<}_{2n}} (-)^{\pi} P_{\pi(1),\pi(2)}P_{\pi(3),\pi(4)}\dots 
P_{\pi(2n-1),\pi(2n)} = 
$$
\beq\left.
\begin{array}{clllllcl} 
  \left | \right . P_{1,2} & P_{1,3}  & \dots &P_{1,R} & P_{1,R+1} 
& P_{1,R+2} & \dots & P_{1,2R}  \\
           & P_{2,3} & \dots &P_{2,R}&  P_{2,R+1} & P_{2,R+2}& \dots & P_{2,2R}\\
           &        & \phantom{c} . & \phantom{c} .         & \phantom{c} .           & \dots  & \dots & \phantom{c} .          \\
           &        &  & \dots     & \phantom{c} .           & \dots  & \dots & \phantom{c} .          \\
           &        &   &           & P_{R,R+1} &  P_{R,R+2} & \dots &  P_{R,2R}  \\
           &        &   &           &             & P_{R+1,R+2} &\dots &  P_{R+1,2R} \\  
           &        &   &           &             &                     &\dots & \phantom{c} .         \\  
           &        &   &           &             &                     &  & P_{2R-1,2R}   
\end{array} \right | \label{pfaffian-structure}
\eeq
where ${\cal S}^{<}_{2n}$ denotes all elements $\pi$ 
of the symmetric group ${\cal S}_{2n}$ which give {\em ordered pairs}; 
i.e.  $\pi(2l-1)<\pi(2l)$) and $\pi(2l-1)<\pi(2m-1)$ for $l<m$.
We particularly make use of the known property that a Pfaffian can be expanded 
along ``rows'' or ``columns'', where the $r$-th row or column 
corresponds to all $P_{i,j}$ with $i=r$ or $j=r$.
In analogy to matrix minors, we will call the {\em minor Pfaffian} 
$\hat{P}_{i,j}\equiv\hat{P}_{j,i}$
the Pfaffian of the above structure (\ref{pfaffian-structure}) when having canceled
the $i$-th and $j$-th row. In terms of these minors the expansion reads
\beq\label{pfaff-expansion}
{\cal P}_{2R}=\sum_{i=1\atop i\neq r}^{2R-1} 
	(-)^{i+r+1} P_{\overrightarrow{i,r}} \hat{P}_{\overrightarrow{i,r}}\, ,
\eeq
where $\overrightarrow{i,r}$ means that the indices are to be written in increasing order.
It is worth noting that the r-th part of this expansion reflects 
all possible contractions with the field operator $\Psi_r$ 
performed in Eq.(\ref{vacuum-expect}).

There are two cases which we will study in this work: $\ket{0}$
being $(i)$ the ground state, denoted by $\ket{g}$ and 
      $(ii)$ the $c$-vacuum, denoted by $\ket{\Downarrow}$.

\subsection{Ground state}

At equilibrium~\cite{Lieb61,Pfeuty70,Mccoy70}, 
a crucial simplification is that 
$\langle A_l A_m \rangle_{g}=-\langle B_l B_m \rangle_{g}=\delta_{lm}$. 
This reduces the Pfaffian to a T\"oplitz determinant
$$
 \left .
\begin{array}{clllllcll} 
\langle S_l^\alpha S_{l+R}^\alpha \rangle_{g}= {s(\alpha,\alpha)}  \left | \right . 0 & \dots   & 0 & G^{\alpha\alpha}_{1,1}
& G^{\alpha\alpha}_{1,2}   &  \dots & G^{\alpha\alpha}_{1,R}  \\
           & \dots  &\dots &  \dots & \dots   & \dots & \dots \\
           &        & 0          & G^{\alpha\alpha}_{R-1,1} &  G^{\alpha\alpha}_{R-1,2} & \dots &  G^{\alpha\alpha}_{R-1,R}  \\
           &        &           & G^{\alpha\alpha}_{R,1} &  G^{\alpha\alpha}_{R,2} & \dots &  G^{\alpha\alpha}_{R,R}  \\
           &        &           &             & 0 &\dots &  0 \\  
           &        &           &                                  & &\dots & \dots         \\  
           &        &           &                               &   &  & 0  

\end{array}
\right |
$$
\beq \label{toeplitz}
=(-)^{R (R-1)/2} s(\alpha,\alpha) \left | 
\begin{array}{lll}  
G^{\alpha\alpha}_{1,1} &  \dots & G^{\alpha\alpha}_{1,R} \\
 \dots  & \dots & \dots \\ 
G^{\alpha\alpha}_{R,1} & \dots &  G^{\alpha\alpha}_{R,R}
\end{array}
\right|
\eeq
with~\cite{Mccoynotation}
\beqa
G^{xx}_{\mu,\nu}&=& \langle A_{l+\mu} B_{l+\nu-1}\rangle_{g} \\ 
G^{yy}_{\mu,\nu}&=& \langle A_{l+\mu-1} B_{l+\nu}\rangle_{g} 
\eeqa 
$\expect{A_l B_m}_{g}={\frak Z}_{m-l}$.
The correlation functions 
$\langle S_l^x S_{l+R}^y \rangle_{g}$ and $\langle S_l^y S_{l+R}^x \rangle_{g}$
identically vanish, since a complete row and column in 
the corresponding matrices vanishes, respectively.
It is worth noting that due to the translational invariance of the state
the determinant above is of T\"oplitz type. As a consequence 
the asymptotics of the correlation functions  
can be extracted explicitly~\cite{Mccoy70} applying the Szeg\"o theorem.

\section{Correlation functions out of equilibrium}\label{result}

As already mentioned, time-dependent correlation functions were derived 
in Ref.~\cite{Mccoy70}. There, the time-dependence was explicitly 
induced into the Hamiltonian (time dependent external magnetic field).
In contrast, we compute matrix elements of operators at non-equilibrium,
meaning that the initial and final state are not eigenstates of the Hamiltonian;
the resulting quantities are then time dependent although the Hamiltonian 
is not. 
First, we consider matrix elements in the $c$-vacuum $\ket{\Downarrow}$ 
(for generic $\gamma$ this is not an eigenstate of the Hamiltonian; 
only for $\gamma=0$ it coincides with the ground state),  
and in excitations on it and on the ground state. 

\subsection{Correlations in the $c$-Vacuum}

Using Eqs.~(\ref{A-of-t},\ref{B-of-t}), leads to the following contractions 
as building blocks for the Pfaffians
$$
\expect{A_l(t) B_m(t)}_{\Downarrow}=\sum_j \mu^*_{j-l}\mu_{m-j}
=
$$
\beq
=\delta_{lm} - 4 \bigfrac{1}{L}\sum_q \left(2 \alpha_q^2\beta_q^2 
	\cos q(m-l) 
	+ \alpha_q\beta_q (1-2 \beta^2_q) \sin q(m-l)\right)
	\sin^2{\Lambda_q t} 
\eeq
\beq
\expect{A_l(t) A_m(t)}_{\Downarrow}=\sum_j \mu^*_{j-l}\mu_{j-m}
=\delta_{lm} - 2\i \bigfrac{1}{L}\sum_q \alpha_q\beta_q 
	\sin q(m-l)\sin{2\Lambda_q t}
\eeq
\beq
\expect{B_l(t) B_m(t)}_{\Downarrow}=\sum_j \mu^*_{j-l}\mu_{l-j} 
= -\delta_{lm} - 2\i \bigfrac{1}{L}\sum_q \alpha_q\beta_q 
	\sin q(m-l)\sin{2\Lambda_q t}
\eeq

We are now ready to write down the two-point spin correlation functions,
applying the results from the previous section

$$
\langle S_l^\alpha S_{l+R}^\beta \rangle_{\Downarrow}= {s(\alpha,\beta)}\cdot
$$
\beq
\left .
\begin{array}{clllllcll} 
\cdot \left | \right . I^{\alpha\beta}_{1,2} & \dots  & I^{\alpha\beta}_{1,R-1}  &J^{\alpha\beta}_{1} & F^{\alpha\beta}_{1} 
& G^{\alpha\beta}_{1,2} &\phantom{c} .  &  \dots & G^{\alpha\beta}_{1,R}  \\
           & \dots  & \dots &\dots &  \dots & \dots  & \phantom{c} . & \dots & \dots \\
           &        & I^{\alpha\beta}_{R-2,R-1}  & J^{\alpha\beta}_{R-2}         & F^{\alpha\beta}_{R-2}         & G^{\alpha\beta}_{R-2,2} & \phantom{c} .  & \dots & G^{\alpha\beta}_{R-2,R}         \\
           &        &   & J^{\alpha\beta}_{R-1}          & F^{\alpha\beta}_{R-1} &  G^{\alpha\beta}_{R-1,2} & \phantom{c} .& \dots &  G^{\alpha\beta}_{R-1,R}  \\
           &        &   &           & E^{\alpha\beta} &  D^{\alpha\beta}_{2} & \phantom{c} . & \dots &  D^{\alpha\beta}_{R}  \\
           &        &   &           &             & K^{\alpha\beta}_{2} & \phantom{c} .&\dots &  K^{\alpha\beta}_{R} \\  
           &        &   &           &             &                       & H^{\alpha\beta}_{2,3}  &\dots &  H^{\alpha\beta}_{2,R} \\  
           &        &   &           &             &                     & &\dots & \dots         \\  
           &        &   &           &             &                  &   &  & H^{\alpha\beta}_{R-1,R}  

\end{array}
\right |
\label{pfaffian}
\eeq
where $s(x,x)=s(y,y)=1/4 (-)^{R(R+1)/2}$, 
\beqb{c}{xx}\left.
\begin{array}{ccc}
I^{xx}_{\mu,\nu}&=&  \langle A_{l+\mu}(t) A_{l+\nu}(t)\rangle_{\Downarrow}  \\ \\
J^{xx}_{\mu}  &=&I^{xx}_{\mu,R} \\ \\ 
H^{xx}_{\mu,\nu}&=& \langle B_{l+\mu-1}(t) B_{l+\nu-1}(t)\rangle_{\Downarrow}  \\ \\
K^{xx}_{\nu}&=& H^{xx}_{1,\nu}  \\ \\ 
G^{xx}_{\mu,\nu}&=& \langle A_{l+\mu}(t) B_{l+\nu-1}(t)\rangle_{\Downarrow} \\ \\ 
F^{xx}_{\mu}&=&G^{xx}_{\mu,1}   \\ \\
E^{xx}&=&G^{xx}_{R,1}   \\ \\
D^{xx}_{\nu}&=&G^{xx}_{R,\nu}   
\end{array}\right\}
\eeqb

\beqb{c}{yy}\left.
\begin{array}{ccc}
 I^{yy}_{\mu,\nu}&=&  \langle A_{l+\mu-1}(t) A_{l+\nu-1}(t)\rangle_{\Downarrow}  \\ \\
J^{yy}_{\mu}  &=&I^{yy}_{\mu,R}\\ \\ 
H^{yy}_{\mu,\nu}&=& \langle B_{l+\mu}(t) B_{l+\nu}(t)\rangle_{\Downarrow}  \\ \\
K^{yy}_{\nu}&=& H^{yy}_{1,\nu}  \\ \\ 
G^{yy}_{\mu,\nu}&=& \langle A_{l+\mu-1}(t) B_{l+\nu}(t)\rangle_{\Downarrow} \\ \\ 
F^{yy}_{\mu}&=&G^{yy}_{\mu,1}   \\ \\
E^{yy}&=&G^{yy}_{R,1}   \\ \\
D^{yy}_{\nu}&=&G^{yy}_{R,\nu}   
\end{array}\right\}
\eeqb
and $s(x,y)=s(y,x)= -\i/4 (-)^{R(R-1)/2}$, 
\beqb{c}{xy}\left.
\begin{array}{ccc}
I^{xy}_{\mu,\nu}&=&  \langle A_{l+\mu}(t) A_{l+\nu}(t)\rangle_{\Downarrow}  \\ \\
G^{xy}_{\mu,\nu}&=& \langle A_{l+\mu}(t) B_{l+\nu}(t)\rangle_{\Downarrow} \\ \\
J^{xy}_{\mu}  &=& G^{xy}_{\mu,0} \\ \\ 
F^{xy}_{\mu}&=&G^{xy}_{\mu,1}   \\ \\
H^{xy}_{\mu,\nu}&=& \langle B_{l+\mu}(t) B_{l+\nu}(t)\rangle_{\Downarrow}  \\ \\
E^{xy}&=&H^{xy}_{0,1}   \\ \\
D^{xy}_{\nu}&=& H^{xy}_{0,\nu}  \\ \\ 
K^{xy}_{\nu}&=&H^{xy}_{1,\nu}   
\end{array}\right\}
\eeqb

\beqb{c}{yx}\left.
\begin{array}{ccc}
I^{yx}_{\mu,\nu}&=&  \langle A_{l+\mu-1}(t) A_{l+\nu-1}(t)\rangle_{\Downarrow}  \\ \\
G^{yx}_{\mu,\nu}&=& \langle A_{l+\mu-1}(t) B_{l+\nu-1}(t)\rangle_{\Downarrow} \\ \\
J^{yx}_{\mu}  &=& I^{yx}_{\mu,R} \\ \\ 
F^{yx}_{\mu}&=&I^{yx}_{\mu,R+1}   \\ \\
E^{yx}&=&I^{yx}_{R,R+1}   \\ \\
D^{yx}_{\nu}&=& G^{yx}_{R,\nu}  \\ \\ 
K^{yx}_{\nu}&=&G^{yx}_{R+1,\nu}   \\ \\ 
H^{yx}_{\mu,\nu}&=& \langle B_{l+\mu-1}(t) B_{l+\nu-1}(t)\rangle_{\Downarrow}  
\end{array}\right\}
\eeqb
We note that a Pfaffian $P$ can be written (up to a sign) as
a determinant of the corresponding antisymmetric matrix $A$ of dimension 
$2R\times 2R$~\cite{Muir60} by $[pf P]^2=\det A$. 
Since the $c$-vacuum is transactional invariant 
this determinant is again of T\"oplitz type. Therefore, also here 
the asymptotics of the correlation functions  
could be extracted explicitly along the lines depicted in~\cite{Mccoy70}.

\subsection{Matrix elements for excitations of the vacuum}
\label{app:nonvac-pfaffians}
We now want to concentrate on expectation values for states which 
are not the vacuum.
So let ${\cal C}^\dagger$ and ${\cal C'}^\dagger$ be linear functionals 
in the creation and annihilation operators and let us calculate
\beq\label{non-vacuum-expect}
\bra{{\cal C}}\Psi_1\cdots\Psi_{2R}\ket{{\cal C}'}:=
\bra{0}{\cal C}\Psi_1\cdots\Psi_{2R}{\cal C'}^\dagger\ket{0}
\eeq
Performing all possible contractions in (\ref{non-vacuum-expect}), we obtain
$\bra{0}{\cal C}{\cal C'}^\dagger\ket{0} \bra{0}\Psi_1\cdots\Psi_{2R}\ket{0}$
$+$ all possible contractions where ${\cal C}$ and ${\cal C'}^\dagger$ are
contracted with a pair of field operators, say $\Psi_i$ and $\Psi_j$.
Thus we have to calculate
\beq
\tilde{P}_{i,j}:=\bra{{\cal C}}\Psi_i\ket{0}\bra{0}\Psi_j\ket{{\cal C}'}-
		\bra{{\cal C}}\Psi_j\ket{0}\bra{0}\Psi_i\ket{{\cal C}'},
\eeq
where we take $i<j$ in order to avoid double counting of contractions.
The sign coming from transporting the operators ${\cal C}$ and 
${\cal C'}^\dagger$ to the left of $\Psi_i$ and the right 
of $\Psi_j$ respectively is $(-)^{i+j+1}$. 
In the remaining vacuum expectation value the field operators
$\Psi_i$ and $\Psi_j$ are missing, which corresponds to canceling 
the rows $i$ and $j$ in the original Pfaffian (\ref{pfaffian-structure}). 
Consequently, this expectation value is the 
minor Pfaffian $\hat{P}_{\overrightarrow{i,j}}$, and we obtain
$$
\bra{{\cal C}}\Psi_1\cdots\Psi_{2R}\ket{{\cal C}'}:={\cal P}_{2R}+\sum_{i=1}^{2R-1}
\sum_{j=i+1}^{2R} (-)^{i+j+1} \tilde{P}_{i,j}\hat{P}_{\overrightarrow{i,j}}
= 
$$
$$
={\cal P}_{2R}+\sum_{i=1}^{2R-1}\left(
	\sum_{j=1}^{i-1} (-)^{i+j+1}\/ 0 \cdot\hat{P}_{\overrightarrow{i,j}}
	+\sum_{j=i+1}^{2R} (-)^{i+j+1} \tilde{P}_{i,j}\hat{P}_{\overrightarrow{i,j}}\right)
$$
We note that this expression 
is the sum over Pfaffian expansions (\ref{pfaff-expansion}).
Indeed, each element of the first sum is the expansion of a Pfaffian
along the $i$-th row, in which $P_{j,i}=0$ and 
$P_{i,j}\rightarrow\tilde{P}_{i,j}$, hence
\beq\label{sumofpfaffs}
\bra{{\cal C}}\Psi_1\cdots\Psi_{2R}\ket{{\cal C}'}:=\sum_{i=0}^{2R-1}
 {\cal P}_{2R}^{(i)}\; ,
\eeq
where we defined
$$
 {\cal P}_{2R}^{(0)}:= {\cal P}_{2R}
$$
and
\beq
\left.\begin{array}{ccccccc} \displaystyle {\cal P}_{2R}^{(i)}:=
 \left | \right . 
P_{1,2} &\dots & 0    &P_{1,i+1}        &\dots & P_{1,2R}  \\
        &\dots &\dots &\dots            &\dots & \dots    \\
        &      & 0    &P_{i-1,i+1}      &\dots & P_{i-1,2R}  \\
        &      &      &\tilde{P}_{i,i+1}&\dots &\tilde{P}_{i,2R} \\
        &      &      &                 &\dots & \dots   \\
        &      &      &                 &      &  P_{2R-1,2R}
\end{array} \right | \label{nonvac-pfaffian}
\eeq

Actually, we found the correlations 
$\langle {\cal C} | S_l^\alpha S_m^\beta| {\cal C'}  \rangle$ 
expressed as a {\it sum of}\/ Pfaffians.
The generalization to operators
${\cal C}^\dagger$ that are multi-linear in the annihilation and 
creation operators can be related to a sum of multi-row expanded
Pfaffians\cite{FUTUREWORK}.

In what follows, we will 
come back to the cases of the ground state and the $c$-vacuum, 
discussed in the previous section.
As mentioned, in order to be able to
explicitly extract the asymptotics of the correlations, the
initial and final state have to be transactional invariant.
In the following, the translational invariance is explicitly broken.

\subsubsection{Single hole excitations on the ground state} 

We choose the final and initial state to be
$\bra{{\cal C}}=\bra{g}c^\dagger_j/\sqrt{{\frak B}_0}=:\bra{j}$ and  
$\ket{{\cal C'}}=c^{}_k\ket{g}/\sqrt{{\frak B}_0}=:\ket{k}$,
which is normalized due to
$
\bra{g}c^\dagger_j c^{}_k\ket{g}={\frak B}_{k-j}\; .
$
Then we have to calculate
$$
\expect{A_l B_m}^{jk}_{g}\doteq 
	\bra{j} A_l \ket{g} \bra{g} B_m\ket{k} - 
	\bra{j} B_m \ket{g} \bra{g} A_l\ket{k}
$$
In this case we find
\begin{eqnarray}
\expect{A_l B_m}^{jk}_{g} &=& \bigfrac{{\frak BZ}^{}_{j-l}{\frak BZ}^*_{m-k} 
		+ {\frak BZ}^{}_{m-j}{\frak BZ}^*_{k-l}}{{\frak B}_0}\\
\expect{A_l A_m}^{jk}_{g} &=& \bigfrac{{\frak BZ}^{*}_{k-m}{\frak BZ}^{}_{j-l} 
		- {\frak BZ}^{*}_{k-l}{\frak BZ}^{}_{j-m}}{{\frak B}_0} \\
\expect{B_l B_m}^{jk}_{g} &=& - \bigfrac{{\frak BZ}^{*}_{m-k}{\frak BZ}^{}_{l-j} 
		- {\frak BZ}^{*}_{l-k}{\frak BZ}^{}_{m-j}}{{\frak B}_0} \; ,
\end{eqnarray}
The contraction of ${\cal C}$ with ${\cal C'}$ is
\beq
\bra{0}{\cal C}{\cal C'}\ket{0}=\bigfrac{{\frak B}_{k-j}}{{\frak B}_0}
\eeq

We now discuss the correlation functions  $\expect{S_l^\alpha S_{l+R}^\alpha}$.
The only non-zero contributions come from contractions of ${\cal C}$ and ${\cal C'}$
with one operator of type $A$ and one of type $B$ (since only vacuum expectations
of an equal number of $A$'s and $B$'s are non-zero as discussed before).
That means that here the sum of Pfaffians simplifies to the following 
sum of determinants
$$
\expect{S_l^\alpha S_l^\alpha}=\sum_{i=1}^{R}
{\cal D}_{R}^{(i)}\; 
$$
with 
\beq \label{Sxx:GS}
{\cal D}_{R}^{(i)}=(-1)^{R} \left | 
\begin{array}{lll}  
G^{\alpha\alpha}_{1,1} &  \dots & G^{\alpha\alpha}_{1,R} \\
 \dots  & \dots & \dots \\ 
\tilde{G}^{\alpha\alpha}_{i,1} & \dots &  \tilde{G}^{\alpha\alpha}_{i,R} \\
\dots & \dots &  \dots \\
G^{\alpha\alpha}_{R,1} & \dots &  G^{\alpha\alpha}_{R,R} \\
\end{array}
\right|
\eeq
with
\beqa
G^{xx}_{\mu,\nu}&=& \langle A_{l+\mu} B_{l+\nu-1}\rangle_{g} \\ 
\tilde{G}^{xx}_{\mu,\nu}&=& \expect{A_{l+\mu} B_{l+\nu-1}}^{jk}_{g} \\
G^{yy}_{\mu,\nu}&=& \langle A_{l+\mu-1} B_{l+\nu}\rangle_{g} \\
\tilde{G}^{yy}_{\mu,\nu}&=& \expect{A_{l+\mu-1} B_{l+\nu}}^{jk}_{g}
\eeqa 
Analogously, for the correlation functions  $\expect{S_l^x S_{l+R}^y}$ and 
$\expect{S_l^y S_{l+R}^x}$
the only non-zero contributions come from contractions of ${\cal C}$ and ${\cal C'}$
with two operators of type $B$ and two of type $A$, respectively.
This again simplifies the sum of Pfaffians to a sum of determinants
$$
\expect{S_l^x S_l^y}=\sum_{i=1}^{R}
{\cal P}_{2R}^{(i)}\; , 
$$
where
\beq
\left .
\begin{array}{cllllllclll} 
{\cal P}_{2R}^{(i)}=\left | \right . 0 & \dots  & 0  & G^{xy}_{1,1} & \phantom{c} . & \dots
& 0& G^{xy}_{1,i+1} &\phantom{c} . &  \dots & G^{xy}_{1,R+1}  \\
           & \dots  & \phantom{c} . &\dots & \phantom{c} . & \dots & \phantom{c} . & \phantom{c} . & \phantom{c} . & \dots & \phantom{c} . \\
           &        & 0  & G^{xy}_{R-2,1} & \phantom{c} .  & \dots       & \phantom{c} . & G^{xy}_{R-2,i+1} & \phantom{c} .  & \dots & \phantom{c} . \\
           &        &   & G^{xy}_{R-1,1}  & \phantom{c} .   & \dots     & 0& G^{xy}_{R-1,i+1} &  \phantom{c} .  & \dots & G^{xy}_{R-1,R+1}         \\
           &        &   &      &   0      & \dots   & 0& 0& \phantom{c} .  & \dots &  0  \\
           &        &   &      &          & \dots   & \phantom{c} .& \phantom{c} .&  \phantom{c} . & \dots &   \phantom{c} .   \\
           &        &   &      &          &      & 0& 0&  \phantom{c} .&\dots &  0 \\  
           &        &   &      &          &   &  & \tilde{H}^{xy}_{i,i+1}  &  \phantom{c} .&\dots &  \tilde{H}^{xy}_{i,R+1} \\  
           &        &   &      &          &   &  & & 0 &\dots          & 0      \\  
           &        &   &      &          &   &  & &  &\dots       & \phantom{c} .   \\  
           &        &   &      &          &   &  & & &               & 0  

\end{array}
\right |
\label{pfaffian:GS-xy}
\eeq
and this simplifies to
$$
\expect{S_l^x S_l^y}=\sum_{i=1}^{R}
{\cal D}_{R}^{(i)}\; 
$$
with 
\beq \label{Sxy:GS}
{\cal D}_{R}^{(i)}=(-1)^{R} \left | 
\begin{array}{llllll}  
G^{xy}_{1,1} &  \dots & G^{xy}_{1,i-1} & G^{xy}_{1,i+1} &  \dots & G^{xy}_{1,R+1} \\
 \dots  & \dots & \dots& \dots & \dots  \\ 
G^{xy}_{R-1,1} &  \dots & G^{xy}_{R-1,i-1} & G^{xy}_{R-1,i+1} &  \dots & G^{xy}_{R-1,R+1} \\
0 & \dots & 0 & \tilde{H}^{xy}_{i,i+1} & \dots & \tilde{H}^{xy}_{i,R+1} 
\end{array}
\right|
\eeq
with
\beqa
G^{xy}_{\mu,\nu}&=& \langle A_{l+\mu} B_{l+\nu-1}\rangle_{g} \\ 
\tilde{H}^{xy}_{\mu,\nu}&=& \expect{B_{l+\mu-1} B_{l+\nu-1}}^{jk}_{g}
\eeqa 
In an analogous way we find
$$
\expect{S_l^y S_l^x}=\sum_{i=1}^{R}
{\cal D}_{R}^{(i)}\; 
$$
with ${\cal D}_{R}^{(i)}$ defined as in (\ref{Sxy:GS}), but
\beqa
G^{xy}_{\mu,\nu}&=& \langle A_{l+\nu-1} B_{l+\mu}\rangle_{g} \\ 
\tilde{H}^{xy}_{\mu,\nu}&=& \expect{A_{l+\mu-1} A_{l+\nu-1}}^{jk}_{g}
\eeqa

\subsubsection{Single particle excitations on the ground state}

Alternatively we consider
$\bra{{\cal C}}=\bra{g}c^{}_j/\sqrt{{\frak A}_0}=:\bra{j}$ and  
$\ket{{\cal C'}}=c^{\dagger}_k\ket{g}/\sqrt{{\frak A}_0}=:\ket{k}$ as
final and initial state, which are again normalized according to
$
\bra{g}c^\dagger_j c^{}_k\ket{g}={\frak A}_{k-j}\; .
$
In this case the possible contractions with the operators ${\cal C}$ 
and ${\cal C'}$ are
\begin{eqnarray}
\expect{A_l B_m}^{jk}_{g} &=& -\bigfrac{{\frak AZ}^{}_{k-l}{\frak AZ}^*_{m-j} 
		+ {\frak AZ}^{}_{m-k}{\frak AZ}^*_{j-l}}{{\frak A}_0}\\
\expect{A_l A_m}^{jk}_{g} &=& \bigfrac{{\frak AZ}^{*}_{j-m}{\frak AZ}^{}_{k-l} 
		- {\frak AZ}^{*}_{j-l}{\frak AZ}^{}_{k-m}}{{\frak A}_0} \\
\expect{B_l B_m}^{jk}_{g} &=& - \bigfrac{{\frak AZ}^{*}_{m-j}{\frak AZ}^{}_{l-k} 
		- {\frak AZ}^{*}_{l-j}{\frak AZ}^{}_{m-k}}{{\frak A}_0} \; ,
\end{eqnarray}
The contraction of ${\cal C}$ with ${\cal C'}$ is here
\beq
\bra{0}{\cal C}{\cal C'}\ket{0}=\bigfrac{{\frak A}_{k-j}}{{\frak A}_0}
\eeq

\subsubsection{Single particle excitations on the $c$-vacuum} 

We take the final and initial state to be $\bra{{\cal C}}=\bra{0}c_j=:\bra{j}$ and  
$\ket{{\cal C'}}=c_{k}^\dagger=:\ket{k}$.
In this case the Pfaffian ${\cal P}_{2R}^{i}$ is given by Eqs.~(\ref{pfaffian})--(\ref{yx}),
where (following Eq.~\ref{nonvac-pfaffian}) in the $i$-th row 
$$
\bra{\Downarrow}A_l B_m\ket{\Downarrow} \longrightarrow 
\expect{A_l B_m}_\Downarrow^{jk} 
$$
are replaced with 
$$
\expect{A_l B_m}_\Downarrow^{jk}\doteq 
\bra{j} A_l \ket{\Downarrow} \bra{\Downarrow} B_m\ket{k} - 
	\bra{j} B_m \ket{\Downarrow} \bra{\Downarrow} A_l\ket{k}
$$
and in the same manner 
$
\bra{\Downarrow}A_l A_m\ket{\Downarrow} \longrightarrow 
\expect{A_l A_m}_\Downarrow^{jk} 
$, 
$
\bra{\Downarrow}B_l B_m\ket{\Downarrow} \longrightarrow 
\expect{B_l B_m}_\Downarrow^{jk} 
$. 
We find
\begin{eqnarray}
\expect{A_l(t) B_{m}(t)}_\Downarrow^{jk} &=& - \left( \mu_{m-j} \mu^*_{k-l}+ 
					   \mu_{j-l} \mu^*_{m-k}\right)\\
\expect{A_l(t) A_{m}(t)}_\Downarrow^{jk} &=& \mu_{j-l} \mu^*_{k-m} - 
					   \mu_{j-m} \mu^*_{k-l} \\
\expect{B_l(t) B_{m}(t)}_\Downarrow^{jk} &=&  \mu_{m-j} \mu^*_{l-k} - 
					   \mu_{l-j} \mu^*_{m-k}\; .
\label{pfaffian-elements}
\end{eqnarray}
With these results, 
all spin-correlation functions can be calculated as long as
$\bra{{\cal C}} S_l^{x}(t=0) \ket{{\cal C}'} =
\bra{{\cal C}} S_l^{y}(t=0) \ket{{\cal C}'} =0$. In this case it will remain 
zero during the evolution.  This is satisfied if
the parity symmetry of the Hamiltonian is not broken by neither the initial 
nor the final state.

\section{Conclusions}

We calculated exactly 
spin-spin correlations out of equilibrium. For excitations on the ``vacuum'',
they can be written as a sum of Pfaffians.
For excitations on the ground state these Pfaffians reduce to determinants
(see Eqs. (\ref{sumofpfaffs}), (\ref{nonvac-pfaffian})).
The result for particle and hole excitations on the ground state 
and the $c$-vacuum were based on different approaches, writing the
fermionic field operators in the Schr\"odinger and Heisenberg picture,
respectively. 
Comparing with the known eigenstate correlations, we remark that here 
$ \langle S_l^\alpha S_{l+R}^\alpha \rangle$  cannot be reduced  
to $R\times R $ T\"oplitz determinants. 
For the vacuum-correlation functions the Pfaffians instead, they can be related 
to $2R\times 2 R$ T\"oplitz determinants. For correlation functions in excited states 
(of the vacuum) the translational invariance of the system 
is explicitly broken and then the determinants are not anymore of T\"oplitz type. 
This last issue constitutes a further difficulty of the problem to find the 
asymptotics of the correlations since the Szeg\"o theorem cannot be applied.

We have used these results explicitly for studying the dynamics of 
correlations and quantum information theoretic quantities like
the entanglement in specific states\cite{DYNAMICS} but
is also a key ingredient for the study of transport properties of the system. 
One possible application of the  exact results we found here is to study 
the  quantum phase transitions (characteristic of this class of models) 
out of equilibrium.

\acknowledgments
The authors would like to thank G. Falci, R. Fazio, and F. Plastina 
for helpful discussions.  
This work was supported by the European Community (IST-SQUIBIT).

%\section*{References}


\begin{thebibliography}{100}
\bibitem{AUERBACH} A. Auerbach, {\em Interacting Electrons and Quantum Magnetism}, (Springer, Berlin, 1998).
\bibitem{KOREPIN-BOOK}V.E. Korepin, N.M. Bogoliubov, and A.G. Itzergin, 
	{\it Quantum Inverse Scattering Method and Correlation Functions},
	(Cambridge Univ. Press, Cambridge 1993).
\bibitem{INVERSION1} N. Kitanine, J.M. Maillet, N.A. Slavnov, and V. Terras,
	Nucl.Phys. B {\bf 641}, 487-518 (2002).
\bibitem{INVERSION2} F. G\"ohmann and V. E. Korepin,
	J.Phys. A {\bf 33}, 1199-1220 (2000). 
\bibitem{Lieb61}
    E. Lieb, T. Schultz, D. Mattis, Ann. Phys. NY {\bf 60}, 407 (1961).
\bibitem{Pfeuty70}
    P. Pfeuty, Ann. Phys. (N.Y.), Ann. Phys.  {\bf 57}, 79-90 (1970).
\bibitem{Mccoy70}
    E. Barouch, B.M. McCoy, and M. Dresden,
    Phys. Rev. A. {\bf 2}, 1075-1092 (1970);
    E. Barouch and B.M. McCoy,
    Phys. Rev. A. {\bf 3}, 786-804 (1971).
\bibitem{ITS}A.R. Its, A.G. Itzergin, V.E. Korepin, and N.A. Slavnov,
    Phys. Rev. Lett. {\bf 70}, 1704 (1993).
\bibitem{Khveshchenko03}
    D. V. Khveshchenko, cond-mat/0301111, (unpublished).
\bibitem{Nielsen00}
    M. Nielsen  and I. Chuang,
    {\it Quantum Computation and Quantum Communication},
    (Cambridge University Press,  Cambridge, 2000).
\bibitem{Preskill00}
    J. Preskill,  J. Mod. Optics {\bf 47}, 127-137 (2000).
\bibitem{Oconnors01}
    K.M. O'Connors  and W.K. Wootters,
    Phys. Rev. A {\bf 63}, 052302 (2001).
\bibitem{Arnesen01}
    M.C. Arnesen, S. Bose, and V. Vedral,
    Phys. Rev. Lett. {\bf 87}, 017901 (2001).
\bibitem{Osterloh02}
    A. Osterloh, L. Amico, G. Falci, and R. Fazio,
    Nature {\bf 416}, 608 (2002).
\bibitem{Osborne02}
    T. J. Osborne and M. A. Nielsen,
    Phys. Rev. A {\bf 66}, 044301 (2002).
\bibitem{Vidal02}
    G. Vidal, J. I. Latorre, E. Rico, A. Kitaev,
    Phys. Rev. Lett.{\bf 90}, 227902 (2003).
\bibitem{Delgado02} 
	M.A. Martin-Delgado, quant-ph/0207026, (unpublished).
\bibitem{Hines02} 
	A. P. Hines, R. H. McKenzie, and G. J. Milburn,
	quant-ph/0209122, (unpublished).
\bibitem{DYNAMICS} L. Amico, A. Osterloh, F. Plastina, R. Fazio, and M. Palma, 
	to be published in Phys. Rev. A; {\it quant-phys/0307048}.
\bibitem{Caianello52} 
	E. R. Caianello and S. Fubini, Nuovo Cimento {\bf 9}, 1218 (1952).
\bibitem{Mccoynotation} The link to the notation used in Ref.~\cite{Mccoy70} 
	is established by noting that we defined ${\frak Z}_{m-l}=\expect{A_l B_m}_{g}$,
	which correspond to the quantity $G_{m-l}$ in Ref.~\cite{Mccoy70}. 
\bibitem{Muir60} T. Muir, {\it Treatise on the theory of determinants}, 
	(Dover, New York 1960).
\bibitem{FUTUREWORK} L. Amico and A. Osterloh, in preparation.
\end{thebibliography}
\end{document}